\documentclass[aps,pra,showpacs,floatfix,twocolumn,superscriptaddress]{revtex4-1}
\usepackage{graphicx}
\usepackage{amsmath}
\usepackage{bm}
\usepackage{color}
\usepackage{dcolumn}
\usepackage{multirow}
\usepackage{xifthen}
\usepackage{hyperref}

\begin{document}

\newcommand{\etal}{{\it et al.}}
\newcommand{\pprime}{{\prime\prime}}


\newcommand{\NIST}{
National Institute of Standards and Technology, 325 Broadway, Boulder, Colorado 80305, USA}
\newcommand{\CU}{
Department of Physics, University of Colorado, 440 UCB, Colorado 80309, USA}

\title{Hyperfine-mediated electric quadrupole shifts in Al$^+$ and In$^+$ ion clocks}

\author{K. Beloy}
\affiliation{\NIST}
\author{D. R. Leibrandt}
\affiliation{\NIST}
\affiliation{\CU}
\author{W. M. Itano}
\affiliation{\NIST}

\date{\today}

\begin{abstract}
We evaluate the electric quadrupole moments of the ${^1}\!S_0$ and ${^3}\!P_0$ clock states of $^{27}$Al$^+$ and $^{115}$In$^+$. To capture all dominant contributions, our analysis extends through third order of perturbation theory and includes hyperfine coupling of the electrons to both the magnetic dipole and electric quadrupole moments of the nucleus. For $^{27}$Al$^+$, a fortuitous cancellation leads to a suppressed frequency shift. This should allow for continued improvement of the clock without special techniques to control or cancel the shift, such as the averaging schemes that are critical to other optical ion clocks.
\end{abstract}


\maketitle

\section{Introduction}
State-of-the-art optical ion clocks owe their low systematic uncertainties, in part, to the external electric fields that provide strong confinement to the ion being probed. Electric field gradients are an inevitable byproduct of this trapping and, when coupled with the electric quadrupole ($E2$) moments of the clock states, induce substantial frequency shifts in clocks based, for example, on Hg$^+$~\cite{RosHumSch08}, Sr$^+$~\cite{MadDubZho12,BarHuaKle14}, and Yb$^+$~\cite{GodNisJon14,HunSanLip16}. Exploiting the rotational symmetry of the $E2$ interaction allows cancellation of the shift by averaging over different magnetic substates or bias magnetic field orientations \cite{Ita00,DubMadBer05}. The additional operational burden of these averaging schemes is accompanied by distinct technical challenges, such as maintaining stringent control over the bias magnetic field direction.

In contrast to the clocks above, the Al$^+$ clock~\cite{RosHumSch08,ChoHumKoe10} employs two $J=0$ clock states, where $J$ specifies the total electronic angular momentum. Given the spherical symmetry implied by $J=0$, the electrons nominally do not contribute to an $E2$ moment. Rather, the $E2$ moments of the clock states reduce to the negligibly small, state-independent nuclear $E2$ moment. To-date, this has allowed the Al$^+$ clock to operate without special regard for electric field gradients, including those attributed to the co-trapped Be$^+$ or Mg$^+$ logic ion.

Under closer scrutiny, the $J=0$ symmetry of the clock states is weakly broken by the coupling of the electrons to the electromagnetic moments of the nucleus, i.e.{} the hyperfine interaction. This gives rise to state-dependent ``hyperfine-mediated'' corrections to the $E2$ moments, which may readily overshadow the bare nuclear $E2$ moment. Estimates of these effects have given justification for their prior neglect~\cite{ItaBerBru07}. However, given the continued improvement of the Al$^+$ clock~\cite{CheBreHum16}, a more refined analysis is now warranted. In this paper, we evaluate the $E2$ moments of the Al$^+$ clock states using methods of many-body atomic structure theory and discuss implications for the Al$^+$ clock going forward. We further extend our analysis to In$^+$, motivated by the pursuit of a many-ion clock based on this species in Braunschweig~\cite{HerPykKel12}.

Hyperfine-mediated $E2$ moments have recently been analyzed for ground-state alkali-metal atoms, with potential relevance to microwave atomic clocks~\cite{Der16}. As one critical difference to that work, here we find it necessary to proceed through third order of perturbation theory (first order in the external field, second order in the hyperfine interaction) to capture all dominant effects.

Gaussian electromagnetic expressions are employed throughout. We let $e$ and $a_B$ denote the elementary charge and the Bohr radius, respectively.

\section{The $E2$ energy shift}

Here we present formulas for the $E2$ energy shift for an atom in a hyperfine state $|nFM\rangle$, where $n$ identifies the electronic state (inclusive of $J$), nuclear spin $I$ is implicit, and $F$ and $M$ specify the total angular momentum and its projection onto the $z$-axis, respectively. To clarify, here $|nFM\rangle$ represents an eigenstate of the full atomic Hamiltonian, inclusive of nuclear spin and the hyperfine interaction. Generally, we are interested in a scenario in which static and radiofrequency electric fields simultaneously perturb the atom. Here we limit our concern to effects first order in the field, in which case only the static component needs to be considered. Describing this field with the electric potential $\Phi$, the $E2$ interaction can be written as the scalar product
\begin{equation}
V_{E2}=\mathcal{E}_2\cdot\mathcal{Q}_2,
\label{Eq:VE2nosplit}
\end{equation}
where $\mathcal{E}_2$ is shorthand for the field-gradient tensor
\begin{equation*}
\mathcal{E}_2\equiv \frac{1}{\sqrt{6}}\left\{\bm{\nabla}\otimes\bm{\nabla}\right\}_2\Phi,
\end{equation*}
understood to be evaluated at the center-of-mass position of the atom. Here $\left\{\bm{\nabla}\otimes\bm{\nabla}\right\}_2$ represents the second rank tensor formed by coupling $\bm{\nabla}$ with itself~\cite{VarMosKhe88}. $\mathcal{Q}_2$ is a rank-2, even-parity operator that acts on both electronic and nuclear coordinates. It is given by $\mathcal{Q}_{2}=\sum_iq_ir_i^2C_{2}\left(\mathbf{\hat{r}}_i\right)$, with $q_i$ the charge and $\mathbf{r}_i$ the position of the $i$-th electron or nucleon ($\mathbf{r}_i=r_i\mathbf{\hat{r}}_i$) and with $C_{2}(\mathbf{\hat{r}})$ being the conventional rank-2 $C$-tensor (normalized spherical harmonic).

The first order $E2$ energy shift to the state $|nFM\rangle$ reads
\begin{equation*}
\delta E
=\langle nFM|V_{E2}|nFM\rangle
=\mathcal{E}_{20}\langle nFM|\mathcal{Q}_{20}|nFM\rangle,
\end{equation*}
where, following from elementary angular selection rules, the matrix element $\langle nFM|\mathcal{Q}_{20}|nFM\rangle$ is non-zero only for $F\geq1$. Generally, $V_{E2}$ mixes states of different $M$, suggesting the need to diagonalize a $(2F+1)$-dimensional matrix. Here contributions off-diagonal in $M$ are assumed to be negligible. In practice this is ensured by application of a sufficiently large bias magnetic field, which defines the $z$-axis and lifts the degeneracy in the $M$ states. This permits us to take $M$ as a ``good'' quantum number and has the consequence of isolating the zero-component of the scalar product in Eq.~(\ref{Eq:VE2nosplit}). The zero-component of $\mathcal{E}_{2}$ is
\begin{equation*}
\mathcal{E}_{20}
=\left(-\frac{1}{6}\frac{\partial^2}{\partial x^2}-\frac{1}{6}\frac{\partial^2}{\partial y^2}+\frac{1}{3}\frac{\partial^2}{\partial z^2}\right)\Phi
=\frac{1}{2}\frac{\partial^2 \Phi}{\partial z^2},
\end{equation*}
where we invoked $\nabla^2\Phi=0$ to arrive at the last expression.

We proceed to define the $E2$ moment $\Theta$ as the expectation value of $\mathcal{Q}_{20}$ for the ``stretched'' state,
\begin{equation*}
\Theta\equiv\langle nFF|\mathcal{Q}_{20}|nFF\rangle.
\end{equation*}
Meanwhile, from the Wigner-Eckart theorem,
\begin{align*}
\langle nFM|\mathcal{Q}_{20}|nFM\rangle
={}&(-1)^{F-M}
\frac{\left(\begin{array}{ccc}
F & 2 & F \\
-M & 0 & M
\end{array}
\right)}
{\left(\begin{array}{ccc}
F & 2 & F \\
-F & 0 & F
\end{array}
\right)}
\\&\times
\langle nFF|\mathcal{Q}_{20}|nFF\rangle.
\end{align*}
Evaluating the 3$j$-symbols and combining the preceding expressions, we arrive at the energy shift expression
\begin{equation*}
\delta E
=\frac{1}{2}\frac{\partial^2 \Phi}{\partial z^2}\frac{3M^2-F(F+1)}{F(2F-1)}
\Theta.
\end{equation*}
Note that the $M$-dependence appears explicitly in this last expression, with $\Theta$ itself being independent of $M$.

The definition of the $E2$ moment $\Theta$ is chosen to be consistent with previous works for $J\geq1$ atomic states, where nuclear structure and hyperfine effects are typically neglected. Expressions in the following section make explicit reference to the {\it nuclear} $E2$ moment $Q$. For this quantity, we adhere to the conventional nuclear physics definition, which incorporates an additional factor of two. Specifically, in a complete absence of electrons, $\Theta$ equates to $Q/2$. In any case, our formulas are self-consistent and unambiguously specify the physical meaning of $\Theta$.

\section{Hyperfine-mediated $E2$ moments for $J=0$ states}

The expressions in the previous section are valid for generic hyperfine states. In the remainder, we limit our attention to $J=0$ atomic states, with the immediate consequence that $F=I$ ($I\geq1$ assumed throughout). To arrive at practical expressions for evaluating the $E2$ moments, we treat the $E2$ interaction and the hyperfine interaction on equal footing as perturbations,
\begin{equation}
\begin{gathered}
V_{E2}=
\mathcal{E}_2\cdot\mathcal{Q}^{(e)}_2
+\mathcal{E}_2\cdot\mathcal{Q}^{(n)}_2,
\\
V_\mathrm{hfi}=\mu^{(n)}_1\cdot\mathcal{T}^{(e)}_1
+\mathcal{Q}^{(n)}_2\cdot\mathcal{T}^{(e)}_2.
\end{gathered}
\label{Eq:Vboth}
\end{equation}
Subscripts appearing on the right-hand-side of the expressions specify tensor rank, while superscripts identify operators acting exclusively on electronic ($e$) or nuclear ($n$) coordinates, inclusive of spin. Note that the operator $\mathcal{Q}_2$ of the preceding section has been partitioned into electronic and nuclear parts. $V_\mathrm{hfi}$ accounts for coupling of the electrons to the magnetic dipole and electric quadrupole moments of the nucleus. We will refer to the respective terms as the $M1$-hyperfine interaction and $E2$-hyperfine interaction, where the latter is not to be confused with the $E2$ interaction with the external field, $V_{E2}$. In the limit of a point nucleus, $\mathcal{T}^{(e)}_1=-e\sum_i(1/r_i^2)\left(\bm{\alpha}_i\times\mathbf{\hat{r}}_i\right)$ and $\mathcal{T}^{(e)}_2=-e\sum_i(1/r_i^3)C_2\left(\mathbf{\hat{r}}_i\right)$, where $\bm{\alpha}$ is composed of the conventional Dirac matrices ($\alpha_x,\alpha_y,\alpha_z$) and the summations run over all electrons. We neglect coupling of the electrons to higher-multipolar moments of the nucleus ($M3$-hyperfine, $E4$-hyperfine, etc.).

The $E2$ moment $\Theta$ is attributed to energy shifts first order in $V_{E2}$ and all orders in $V_\mathrm{hfi}$. In practice, however, we only retain the leading contributions in $V_\mathrm{hfi}$. The notation $\Theta^{(1+m)}$ is used to denote the $(1+m)$-th order contribution, with $m$ being the number of hyperfine interactions involved. Applying conventional perturbation theory through third order, we arrive at the following expressions specific to $J=0$ states,
\begin{gather*}
\Theta^{(1+0)}=\frac{1}{2}Q,
\\
\Theta^{(1+1)}=\frac{1}{5}Q
\left[\mathcal{Q}_2,\mathcal{T}_2\right]_1,
\\
\begin{aligned}
\Theta^{(1+2)}={}&
\frac{2}{\sqrt{15}}\mu^2 A_{11;I}
\left[\mathcal{Q}_2,\mathcal{T}_1,\mathcal{T}_1\right]
\\&
+\frac{1}{3}\mu^2 A_{11;I}
\left[\mathcal{T}_1,\mathcal{Q}_2,\mathcal{T}_1\right]
\\&
+\frac{1}{5}\mu Q A_{12;I}
\left[\mathcal{Q}_2,\mathcal{T}_1,\mathcal{T}_2\right]
\\&
+\frac{1}{\sqrt{15}}\mu Q A_{12;I}
\left[\mathcal{Q}_2,\mathcal{T}_2,\mathcal{T}_1\right]
\\&
-\frac{1}{\sqrt{15}}\mu Q A_{12;I}
\left[\mathcal{T}_1,\mathcal{Q}_2,\mathcal{T}_2\right]
\\&
+\frac{1}{10}Q^2 A_{22;I}
\left[\mathcal{Q}_2,\mathcal{T}_2,\mathcal{T}_2\right]
\\&
+\frac{1}{20}Q^2 A_{22;I}
\left[\mathcal{T}_2,\mathcal{Q}_2,\mathcal{T}_2\right]
\\&
+\frac{1}{6}\mu^2 QB_{1;I}
\left[\mathcal{T}_1,\mathcal{T}_1\right]_2
\\&
+\frac{1}{40}Q^3B_{2;I}
\left[\mathcal{T}_2,\mathcal{T}_2\right]_2,
\end{aligned}
\end{gather*}
where $\mu$ and $Q$ are the conventional nuclear $M1$ and $E2$ moments. The $I$-dependent angular factors $A_{k_1k_2;I}$ and $B_{k;I}$ are given in terms of $3j$- and $6j$-symbols by
\begin{gather*}
A_{k_1k_2;I}=
(-1)^{2I}
\frac{
\left(\begin{array}{ccc}
I & 2 & I \\
-I & 0 & I
\end{array}
\right)
\left\{\begin{array}{ccc}
k_1 & k_2 & 2 \\
I & I & I
\end{array}
\right\}
}{\left(\begin{array}{ccc}
I & k_1 & I \\
-I & 0 & I
\end{array}
\right)
\left(\begin{array}{ccc}
I & k_2 & I \\
-I & 0 & I
\end{array}
\right)
},
\\
B_{k;I}=
(-1)^{2I}\frac{
\left\{\begin{array}{ccc}
I & I & k \\
I & I & 2
\end{array}
\right\}
-
\left\{\begin{array}{ccc}
I & I & k \\
I & I & 0
\end{array}
\right\}
}{\left(\begin{array}{ccc}
I & k & I \\
-I & 0 & I
\end{array}
\right)^2
}.
\end{gather*}
The string of operators appearing in square brackets is compact notation for the pure electronic factors
\begin{gather*}
\left[X_{k},Y_{k}\right]_r
=\sum_{n^\prime}\frac{\langle n||X_{k}||n^\prime\rangle\langle n^\prime||Y_{k}||n\rangle}
{\left(E_{n}-E_{n^\prime}\right)^r},
\\
\begin{aligned}
\hspace{5em}&\hspace{-5em}
\left[X_{k_1},Y_{k_2},Z_{k_3}\right]=
\\
&\sum_{n^\prime n^\pprime}\frac{\langle n||X_{k_1}||n^\prime\rangle\langle n^\prime||Y_{k_2}||n^\pprime\rangle\langle n^\pprime||Z_{k_3}||n\rangle}{\left(E_{n}-E_{n^\prime}\right)\left(E_{n}-E_{n^\pprime}\right)},
\end{aligned}
\end{gather*}
where we have dropped the superscript $(e)$ on the operators involved. Here $|n\rangle$ and $E_n$, with arbitrary number of primes attached to the $n$, denote electronic states and energies. That is, these are eigensolutions to the atomic Hamiltonian in absence of nuclear spin and the hyperfine interaction. Reduced matrix elements and energies are independent of the magnetic substates, and it is understood that the sum-over-states expressions above exclude summation over magnetic quantum numbers. The unprimed $n$ designates the $J=0$ state of concern, while $n^\prime$ and $n^\pprime$ run over all other electronic states (having $J^\prime$ and $J^\pprime$, respectively). Selection rules insist that $J^\prime=k$ for the top expression and $J^\prime=k_1$ and $J^\pprime=k_3$ for the bottom expression. Moreover, since only even-parity operators play a role, all states involved must have identical parity.

The first order $\Theta^{(1+0)}$ represents the $E2$ moment in absence of the hyperfine interaction. In accordance with the Introduction, this reduces to the bare nuclear value, with a factor of $1/2$ bridging the atomic physics and nuclear physics definitions of the $E2$ moment. At second order, a correction $\Theta^{(1+1)}$ arises due to the $E2$-hyperfine interaction. Selection rules preclude the $M1$-hyperfine interaction from having effect at this order, prompting further progression to third order. The expression for $\Theta^{(1+2)}$ includes several terms. The first two terms involve two $M1$-hyperfine interactions, while the remaining terms can be regarded as higher-order in interactions that already contribute to $\Theta^{(1+0)}$ and $\Theta^{(1+1)}$. For the specific problem at hand, we further note that the excited clock state is part of a ${^3\!P_J}$ fine structure manifold. The hyperfine-mediated effects are consequently expected to be dominated by hyperfine-mixing of the $^3\!P_0$ clock state with the neighboring $^3\!P_1$ and $^3\!P_2$ states, with corresponding terms in the sum-over-states expressions entering with small energy denominators. From the preceding arguments, we identify one second order term and two third order terms to be of principal interest,
\begin{gather}
-\frac{1}{5}\frac{Q}{\Delta_{20}}
\langle{^3\!P_0}||\mathcal{Q}_2||{^3\!P_2}\rangle
\langle{^3\!P_2}||\mathcal{T}_2||{^3\!P_0}\rangle,
\label{Eq:Q2-T2}\\
\frac{8\sqrt{2}}{75}
\frac{\mu^2}{\Delta_{20}\Delta_{10}}
\langle{^3\!P_0}||\mathcal{Q}_2||{^3\!P_2}\rangle
\langle{^3\!P_2}||\mathcal{T}_1||{^3\!P_1}\rangle
\langle{^3\!P_1}||\mathcal{T}_1||{^3\!P_0}\rangle,
\label{Eq:Q2-T1-T1}\\
\frac{4}{15}
\sqrt{\frac{2}{15}}
\frac{\mu^2}{\Delta_{10}^2}
\langle{^3\!P_0}||\mathcal{T}_1||{^3\!P_1}\rangle
\langle{^3\!P_1}||\mathcal{Q}_2||{^3\!P_1}\rangle
\langle{^3\!P_1}||\mathcal{T}_1||{^3\!P_0}\rangle,
\label{Eq:T1-Q2-T1}
\end{gather}
with fine structure splittings $\Delta_{10}\equiv\left(E_{^3\!P_1}-E_{^3\!P_0}\right)$ and $\Delta_{20}\equiv\left(E_{^3\!P_2}-E_{^3\!P_0}\right)$. Terms (\ref{Eq:Q2-T1-T1}) and (\ref{Eq:T1-Q2-T1}) are specific to $I=5/2$ ($^{27}$Al$^+$). For other $I$, these terms should be multiplied by an additional factor $(5/8)(2I-1)/I$. For the sake of completeness, below we evaluate all first through third order contributions to the $E2$ moments of both clock states. Terms (\ref{Eq:Q2-T2})--(\ref{Eq:T1-Q2-T1}) are found to be comparable to one another and dominate significantly over all other contributions. Given that the fine structure splittings and nuclear moments are known, the critical task reduces to the accurate determination of the five distinct matrix elements appearing in these three terms.

\section{Numerical Analysis and Results}

To evaluate electronic properties, we employ the method of configuration interaction plus many-body perturbation theory (CI+MBPT)~\cite{DzuFlaKoz96,SavJoh02}. In brief, we start with a Dirac-Hartree-Fock description of the atomic core (nuclear charge plus core electrons). The CI procedure treats the strongly-interacting valence electrons in the presence of the ``frozen'' core. The MBPT extension accounts for additional perturbative effects of the valence electrons on the core. Further details of our CI+MBPT implementation, with only minor modification, can be found in Ref.~\cite{BelDerJoh08}.

The CI+MBPT method has been applied to numerous divalent systems, including heavy systems such as Ra~\cite{DzuGin06}. Al$^+$ is comparatively simple. To best gauge the accuracy of our calculations, in particular the matrix elements appearing in terms (\ref{Eq:Q2-T2})--(\ref{Eq:T1-Q2-T1}), we first identify relevant data in the literature for comparison. Hyperfine intervals for the ${^3}\!P_1$ state of $^{27}$Al$^+$~\cite{ItaBerBru07} and the ${^3}\!P_{1,2}$ states of $^{25}$Mg~\cite{Lur62} have been measured. Mg has similar electronic structure to Al$^+$, justifying its inclusion here. Neglecting higher order effects, the hyperfine intervals may be combined with nuclear moments~\cite{Sto05} to infer diagonal $M1$-hyperfine and $E2$-hyperfine matrix elements, $\langle{^3}\!P_J||\mathcal{T}_1||{^3}\!P_J\rangle$ and $\langle{^3}\!P_J||\mathcal{T}_2||{^3}\!P_J\rangle$, for the respective states. These results are presented in Table~\ref{Tab:exptvstheory} with the label ``inferred, uncorrected.'' To these values, we add second order corrections attributed to hyperfine mixing between the neighboring ${^3}\!P_J$ states. The corrections are evaluated using off-diagonal {\it ab initio} CI+MBPT hyperfine matrix elements, together with the nuclear moments and fine structure intervals~\cite{NIST_ASD}. The results are presented in Table~\ref{Tab:exptvstheory} with the label ``inferred, corrected.'' Finally, these values are compared to the respective diagonal {\it ab initio} CI+MBPT matrix elements, labeled ``CI+MBPT.'' 

\newcommand{\Amel}{\langle{^3}\!P_J||\mathcal{T}_1||{^3}\!P_J\rangle}
\newcommand{\Bmel}{\langle{^3}\!P_J||\mathcal{T}_2||{^3}\!P_J\rangle}
\newcommand{\fnm}[1][]{\ifthenelse{\equal{#1}{}}{\footnotemark}{\footnotemark[#1]}}
\newlength{\myl}
\settowidth{\myl}{$\Bmel$}

\begin{table*}[t]
\caption{Diagonal $M1$-hyperfine and $E2$-hyperfine matrix elements for ${^3}\!P_J$ states of Mg, Al$^+$, and In$^+$. Uncorrected matrix elements are inferred from hyperfine intervals~\cite{ItaBerBru07,Lur62,LarHan93} and nuclear moments~\cite{Sto05} found in the literature, neglecting higher order effects of the hyperfine interaction. These results are subsequently corrected for second order effects using off-diagonal {\it ab initio} CI+MBPT hyperfine matrix elements between the ${^3}\!P_J$ states, together with nuclear moments and fine structure intervals~\cite{NIST_ASD}. In many cases, the corrections do not or barely change the expressed result. The corrected matrix elements are then compared to {\it ab initio} CI+MBPT hyperfine matrix elements. All values are in atomic units~\cite{aunote}.}
\label{Tab:exptvstheory}
\begin{ruledtabular}
\begin{tabular}{cccccc}
\vspace{-3mm}\\
& \multicolumn{2}{c}{Mg,~$3s3p\,{^3}\!P_J$}
& \multicolumn{1}{c}{Al$^+$,~$3s3p\,{^3}\!P_J$}
& \multicolumn{2}{c}{In$^+$,~$5s5p\,{^3}\!P_J$}
\\
\cline{2-3}\cline{4-4}\cline{5-6}
\vspace{-3mm}\\
& $J=1$ & $J=2$ & $J=1$ & $J=1$ & $J=2$ \\
\hline
\vspace{-3mm}\\
\multirow{3}{*}{\raggedright\makebox[\myl]{$\Amel$}~$\left\{\begin{array}{l}
\text{inferred, uncorrected}\\
\text{inferred, corrected}\\
\text{CI+MBPT}\\
\end{array}\right.$}	
& $0.07938(<\!1)$	& $0.1573(<\!1)$	& $0.1723(<\!1)$	& $1.059(<\!1)$	& $1.733(4)$ \\
& $0.07936(<\!1)$	& $0.1572(<\!1)$	& $0.1723(<\!1)$	& $1.059(<\!1)$	& $1.733(4)$ \\
& $0.07847$			& $0.1561$			& $0.1696$			& $1.150$		& $1.841$	 \\
\vspace{0mm}\\
\multirow{3}{*}{\raggedright\makebox[\myl]{$\Bmel$}~$\left\{\begin{array}{l}
\text{inferred, uncorrected}\\
\text{inferred, corrected}\\
\text{CI+MBPT}\\
\end{array}\right.$}	
& $-0.482(7)$	& $0.70(1)$	& $-1.72(1)$	& $-6.9(6)$	& $7(2)$	\\
& $-0.466(7)$	& $0.71(1)$	& $-1.26(2)$	& $-6.4(6)$	& $7(2)$	\\
& $-0.4621$		& $0.7023$	& $-1.199$		& $-6.175$	& $8.582$	\\
\vspace{-3mm}\\
\end{tabular}
\end{ruledtabular}
\end{table*}

First, we examine the $M1$-hyperfine matrix elements in Table~\ref{Tab:exptvstheory}. For all three Mg and Al$^+$ states, the second order corrections are small, with the resulting ``inferred, corrected'' values being accurate to within the displayed digits. The corresponding CI+MBPT results exhibit agreement in the range 0.7--1.6\%. This is indicative
of the CI+MBPT accuracy for the $M1$-hyperfine matrix elements between the different ${^3}\!P_J$ states. Next, we examine the $E2$-hyperfine matrix elements for the three Mg and Al$^+$ states. Here, non-negligible uncertainty in the ``inferred, uncorrected'' values stems from uncertainty in the nuclear $E2$ moments. At the same time, the second order corrections are more prominent. These corrections are dominated by terms involving off-diagonal $M1$-hyperfine matrix elements. Our preceding assessment of the CI+MBPT performance for these matrix elements allows us to ascribe a fair uncertainty to the corrections. This is only of significance for the ${^3}\!P_1$ state of Al$^+$, where the correction leads to a doubling of the uncertainty. Finally, for the ${^3}\!P_{1,2}$ states of Mg, the diagonal CI+MBPT matrix elements are found to be within $1\sigma$ ($\sim\!1.5\%$) of the ``inferred, corrected'' values. Meanwhile, for the ${^3}\!P_1$ state of Al$^+$, the CI+MBPT result is found to be within $3\sigma$ ($\sim\!5\%$) of the ``inferred, corrected'' value. This provides us with a measure of the CI+MBPT accuracy for the $E2$-hyperfine matrix elements between the different
${^3}\!P_J$ states. Unfortunately, literature data is lacking to directly assess CI+MBPT performance for $E2$ matrix elements ($\mathcal{Q}_2$ operator) between the ${^3}\!P_J$ states. However, we have no reason to expect the $E2$ matrix elements to be any less accurate than the $M1$-hyperfine or $E2$-hyperfine matrix elements between these states. We will briefly return to the discussion of accuracy below.

Table~\ref{Tab:mels} presents our CI+MBPT results for the five distinct matrix elements appearing in terms (\ref{Eq:Q2-T2})--(\ref{Eq:T1-Q2-T1}). Together with nuclear moments \cite{Sto05} and fine structure splittings \cite{NIST_ASD}, these matrix elements yield the results $-1.08\times10^{-5}\,ea_B^2$ for term (\ref{Eq:Q2-T2}), $3.92\times10^{-6}\,ea_B^2$ for term (\ref{Eq:Q2-T1-T1}), and $5.13\times10^{-6}\,ea_B^2$  for term (\ref{Eq:T1-Q2-T1}). The third order $M1$-hyperfine-mediated terms (\ref{Eq:Q2-T1-T1}) and (\ref{Eq:T1-Q2-T1}) are seen to be comparable in magnitude and add constructively. Together they are close in magnitude to the second order $E2$-hyperfine-mediated term (\ref{Eq:Q2-T2}) but of opposite sign. This results in significant cancellation, with a suppressed cumulative result of $-1.7\times10^{-6}\,ea_B^2$ for these three terms. Clearly, it would have been erroneous to limit our analysis to second order perturbation theory or to account for hyperfine-mediated effects using only the $M1$-hyperfine interaction.

\begin{table}[t]
\caption{{\it Ab initio} CI+MBPT results for the five distinct matrix elements appearing in terms (\ref{Eq:Q2-T2})--(\ref{Eq:T1-Q2-T1}), evaluated for Al$^+$ and In$^+$. State labels refer to the $3s3p\,{^3\!P_J}$ and $5s5p\,{^3\!P_J}$ fine structure manifolds of the respective  ion. All values are in atomic units~\cite{aunote}.}
\label{Tab:mels}
\begin{ruledtabular}
\begin{tabular}{ccc}
& Al$^+$ & In$^+$ \\
\hline
\vspace{-3mm}\\
$\langle{^3\!P_0}||\mathcal{Q}_2||{^3\!P_2}\rangle$	& $-6.271$	& $-6.932$	\\
$\langle{^3\!P_1}||\mathcal{Q}_2||{^3\!P_1}\rangle$	& $-5.428$	& $-5.825$	\\
$\langle{^3\!P_0}||\mathcal{T}_1||{^3\!P_1}\rangle$	& $0.1195$	& $0.7556$	\\
$\langle{^3\!P_1}||\mathcal{T}_1||{^3\!P_2}\rangle$	& $-0.1545$	& $-0.7973$	\\
$\langle{^3\!P_0}||\mathcal{T}_2||{^3\!P_2}\rangle$	& $-1.382$	& $-7.114$
\end{tabular}
\end{ruledtabular}
\end{table}

Despite the cancellation, the terms (\ref{Eq:Q2-T2})--(\ref{Eq:T1-Q2-T1}) provide a correction to the $E2$ moment of the excited clock state three orders of magnitude larger than the bare nuclear value of $2.62\times10^{-9}\,ea_B^2$. Remaining second and third order contributions were evaluated using CI+MBPT matrix elements and energies, with experimental nuclear moments. For both clock states, these contributions amount to $-1\times10^{-8}\,ea_B^2$, verifying the dominance of the terms (\ref{Eq:Q2-T2})--(\ref{Eq:T1-Q2-T1}). Table~\ref{Tab:moments} provides a breakdown of the contributions to the $E2$ moments of both clock states.

\begin{table}[t]
\caption{Contributions to the $E2$ moments of the ground and excited clock states in $^{27}$Al$^+$ and $^{115}$In$^+$. All values are in units of $ea_B^2$. The notation $x[y]$ indicates $x\times10^y$.}
\label{Tab:moments}
\begin{ruledtabular}
\begin{tabular}{lcc}
& $^{27}$Al$^+$ & $^{115}$In$^+$ \\
\hline
&\multicolumn{2}{c}{bare nucleus} \\
$\Theta^{(1+0)}$											& $2.62[-9]$	& $15[-9]$ \\
&\multicolumn{2}{c}{ground clock state} \\
$\Theta^{(1+1)}$											& $-10[-9]$		& \\
$\Theta^{(1+2)}$											& $-0.1[-12]$	& \\
$\Theta$ total												& $-8[-9]$		& \\
&\multicolumn{2}{c}{excited clock state} \\
$\Theta^{(1+1)}$, term (\ref{Eq:Q2-T2})						& $-10.8[-6]$	& $-18.5[-6]$ \\
$\Theta^{(1+1)}$, all other									& $9[-9]$		& \\
$\Theta^{(1+2)}$, term (\ref{Eq:Q2-T1-T1})\footnotemark[1]	& $3.9[-6]$		& $1.1[-6]$ \\
$\Theta^{(1+2)}$, term (\ref{Eq:T1-Q2-T1})\footnotemark[1]	& $5.1[-6]$		& $1.7[-6]$ \\
$\Theta^{(1+2)}$, all other									& $-19[-9]$		& \\
$\Theta$ total												& $-1.7[-6]$	& $-15.7[-6]$ \\
\end{tabular}
\footnotetext[1]{For $^{115}$In$^+$, terms (\ref{Eq:Q2-T1-T1}) and (\ref{Eq:T1-Q2-T1}) are multiplied by $10/9$, as appropriate for the nuclear spin $I=9/2$.}
\end{ruledtabular}
\end{table}

To estimate uncertainty of $\Theta$ for the excited clock state, we ascribe a 3\% uncertainty to each of the five distinct matrix elements entering terms (\ref{Eq:Q2-T2})--(\ref{Eq:T1-Q2-T1}). We assume uncorrelated error in these matrix elements and propagate uncertainty accordingly, rendering a final value $\Theta=-1.7(6)\times10^{-6}\,ea_B^2$.

In Ref.~\cite{ItaBerBru07}, the $E2$ moment of the excited clock state was given as $\Theta\approx-1.2\times10^{-5}\,ea_B^2$, this being an order-of-magnitude larger than the present evaluation. After inspecting notes from that work, a sign error was discovered for the term (\ref{Eq:T1-Q2-T1}) contribution. A large relative error in $\Theta$ resulted from an absence of the cancellation discussed above. Correcting the sign error, the previous result is brought into agreement with the present result.

\section{Implications for the A\lowercase{l}$^+$ ion clock}

Al$^+$ ion clocks to-date have operated with a single Al$^+$ ion and a single logic ion (Be$^+$ or Mg$^+$) simultaneously confined in a linear RF Paul trap \cite{RaiGilBer92}.  For this configuration, there are two contributions to the electric quadrupole field at the position of the Al$^+$ ion: that due to the static axial confining trap potential and that due to the logic ion.  Following Ref.~\cite{WubAmaMan12}, we write the total trap electric potential as
\begin{equation*}
\Phi_T = \frac{V_0}{2} \cos (\Omega t) \frac{X^2 - Y^2}{R^2}+U_0 \frac{Z^2 - \alpha X^2 - (1 - \alpha) Y^2}{d^2},
\end{equation*}
where $V_0/2$ and $\Omega$ are the amplitude and angular frequency of the voltage applied to the rf electrodes, $U_0$ is the voltage applied to the endcap electrodes, $R$ and $d$ are characteristic radial and axial dimensions of the trap, and $\alpha$ parameterizes the radial asymmetry of the static field.  Above, $X$, $Y$, and $Z$ are spatial coordinates in the trap coordinate frame, while $t$ denotes time.  Dropping the rf term, transforming to the quantization coordinate frame, and taking the second derivative with respect to $z$, we get
\begin{equation*}
\frac{\partial^2 \Phi_T}{\partial z^2} = \frac{2 U_0}{d^2} \left[ \frac{3 \cos^2 \theta - 1}{2} - \left( \alpha - \frac{1}{2} \right) \sin^2 \theta \cos(2 \phi) \right],
\end{equation*}
where $\theta$ and $\phi$ are the polar and azimuthal angles of the bias magnetic field as referenced from the trap frame. This result is independent of the position of the Al$^+$ ion in the trap. The electric potential due to the other ion can be written as
\begin{equation*}
\Phi_I = \frac{e}{\sqrt{(X-X_i)^2 + (Y-Y_i)^2 + (Z-Z_i)^2}},
\end{equation*}
where $(X_i, Y_i, Z_i)$ is the position of the logic ion. Transforming to the quantization coordinate frame, taking the second derivative with respect to $z$, and substituting the equilibrium ion positions $(X, Y, Z) = (0, 0, \pm (e d^2/8 U_0)^{1/3})$, we get
\begin{equation*}
\frac{\partial^2 \Phi_I}{\partial z^2} = \frac{2 U_0}{d^2} \frac{3 \cos^2 \theta - 1}{2}.
\end{equation*}
The trap dc potential satisfies
\begin{equation*}
\frac{U_0}{d^2} = \frac{m_1 \omega^2}{2 e} \frac{\mu}{1 + \mu - \sqrt{1 - \mu + \mu^2}},
\end{equation*}
where $m_1$ and $m_2 = \mu m_1$ are the masses of the logic and clock ions and $\omega$ is the angular frequency of the axial common secular mode. Given that $\omega$ is a readily-accessible experimental parameter, $U_0/d^2$ can be immediately determined from the expression above.

For the Al$^+$ clock reported in Ref.~\cite{ChoHumKoe10}, with $\omega/2\pi=3.00$~MHz, $\alpha = 1.65$, and $\theta = \phi = 45^\circ$, we calculate an $E2$ clock shift of $-28$~$\mu$Hz, or $-2.5 \times 10^{-20}$ fractionally. The magnetic field orientation with respect to the trap axes was not characterized with high accuracy in this experiment, but due to geometric constraints we can bound the uncertainty of $\theta$ and $\phi$ to be better than $\pm 5^\circ$. Taking this field-gradient uncertainty together with uncertainty in $\Theta$, we find an uncertainty in the clock shift of $19$~$\mu$Hz, or $1.7\times 10^{-20}$ fractionally. As expected, both the shift and its uncertainty are negligible in comparison with the total clock uncertainty of Ref.~\cite{ChoHumKoe10}. Even as the Al$^+$ clock continues to improve \cite{CheBreHum16}, this shift will likely remain small with respect to the uncertainty budget for the immediate future.

\section{Estimates for an I\lowercase{n}$^+$ many-ion clock}

Following Refs.~\cite{PykHerKel14} and \cite{Kel15}, we consider a many-ion clock based on $^{115}$In$^+$. We evaluate terms~(\ref{Eq:Q2-T2})--(\ref{Eq:T1-Q2-T1}), with appropriate scaling for the nuclear spin $I=9/2$, and neglect all other contributions. Results are tabulated alongside $^{27}$Al$^+$ results in Tables~\ref{Tab:exptvstheory}--\ref{Tab:moments} above. Unlike $^{27}$Al$^+$, there is not significant cancellation among the terms (\ref{Eq:Q2-T2})--(\ref{Eq:T1-Q2-T1}). The second order $E2$-hyperfine-mediated term (\ref{Eq:Q2-T2}) dominates, with a final value of $\Theta=-1.6\times10^{-5}ea_B^2$. Through inspection of Table~\ref{Tab:exptvstheory}, this result is expected to be accurate to within $\sim\!20\%$.

To estimate the clock shift, we assume a linear crystal of eight In$^+$ clock ions and two Yb$^+$ sympathetic cooling ions, with the Yb$^+$ ions residing in the center. We suppose that the trap strength is such that the axial secular frequency of a single $^{172}$Yb$^+$ ion is 330 kHz and that $\alpha\approx1/2$ and $\theta=25^\circ$. By generalizing the above formulas to the many-ion case, we calculate the mean of the $E2$ shift of each of the eight In$^+$ ions to be $-490\,\mu$Hz, or $-3.9\times10^{-19}$ fractionally. The full width of the associated inhomogeneous broadening is $530\,\mu$Hz, which given the lifetime limited transition linewidth of 820 mHz~\cite{BecZanNev01} will not limit the spectroscopy linewidth. We note that at some level it might be necessary to take into account anharmonicity in the trapping potential over the extended size of the ion crystal for a more accurate estimate of the $E2$ shift.

\section{Conclusion}

Here we have derived expressions for hyperfine-mediated electric quadrupole moments of $J=0$ atomic states and have used these expressions to evaluate the corresponding clock shift in optical ion clocks based on $^{27}$Al$^+$ and $^{115}$In$^+$. Interestingly, for $^{27}$Al$^+$, the second order $E2$-hyperfine-mediated contribution is nearly equal but opposite to the third order $M1$-hyperfine-mediated contributions, leading to a substantial suppression of the already small shift.

While the magnitudes of the $E2$ shifts for the clocks considered here are small with respect to current total systematic uncertainties, the $E2$ shift may be more significant for future clocks with lower total systematic uncertainties, clocks with stronger confinement, or clocks based on other atomic species.  For these cases, in addition to the established techniques of cancelling the shift by averaging over different magnetic substates or bias magnetic field orientations, it is worth noting that the shift can also be suppressed by setting $\theta = \arccos(1/\sqrt{3}) \approx 54.7^\circ$ and $\phi = 45^\circ$. Finally, while the quadrupole moments are rather small, it may be possible to do a direct experimental measurement using the technique of Ref.~\cite{ShaAkeOze16}.

\begin{acknowledgements}
The authors thank S.~Brewer and D.~Nicolodi for their careful reading of the manuscript. This work is the contribution of NIST, an agency of the US government, and is not subject to US copyright.
\end{acknowledgements}


\begin{thebibliography}{28}%
\makeatletter
\providecommand \@ifxundefined [1]{%
 \@ifx{#1\undefined}
}%
\providecommand \@ifnum [1]{%
 \ifnum #1\expandafter \@firstoftwo
 \else \expandafter \@secondoftwo
 \fi
}%
\providecommand \@ifx [1]{%
 \ifx #1\expandafter \@firstoftwo
 \else \expandafter \@secondoftwo
 \fi
}%
\providecommand \natexlab [1]{#1}%
\providecommand \enquote  [1]{``#1''}%
\providecommand \bibnamefont  [1]{#1}%
\providecommand \bibfnamefont [1]{#1}%
\providecommand \citenamefont [1]{#1}%
\providecommand \href@noop [0]{\@secondoftwo}%
\providecommand \href [0]{\begingroup \@sanitize@url \@href}%
\providecommand \@href[1]{\@@startlink{#1}\@@href}%
\providecommand \@@href[1]{\endgroup#1\@@endlink}%
\providecommand \@sanitize@url [0]{\catcode `\\12\catcode `\$12\catcode
  `\&12\catcode `\#12\catcode `\^12\catcode `\_12\catcode `\%12\relax}%
\providecommand \@@startlink[1]{}%
\providecommand \@@endlink[0]{}%
\providecommand \url  [0]{\begingroup\@sanitize@url \@url }%
\providecommand \@url [1]{\endgroup\@href {#1}{\urlprefix }}%
\providecommand \urlprefix  [0]{URL }%
\providecommand \Eprint [0]{\href }%
\providecommand \doibase [0]{http://dx.doi.org/}%
\providecommand \selectlanguage [0]{\@gobble}%
\providecommand \bibinfo  [0]{\@secondoftwo}%
\providecommand \bibfield  [0]{\@secondoftwo}%
\providecommand \translation [1]{[#1]}%
\providecommand \BibitemOpen [0]{}%
\providecommand \bibitemStop [0]{}%
\providecommand \bibitemNoStop [0]{.\EOS\space}%
\providecommand \EOS [0]{\spacefactor3000\relax}%
\providecommand \BibitemShut  [1]{\csname bibitem#1\endcsname}%
\let\auto@bib@innerbib\@empty
\bibitem [{\citenamefont {Rosenband}\ \emph {et~al.}(2008)\citenamefont
  {Rosenband}, \citenamefont {Hume}, \citenamefont {Schmidt}, \citenamefont
  {Chou}, \citenamefont {Brusch}, \citenamefont {Lorini}, \citenamefont
  {Oskay}, \citenamefont {Drullinger}, \citenamefont {Fortier}, \citenamefont
  {Stalnaker}, \citenamefont {Diddams}, \citenamefont {Swann}, \citenamefont
  {Newbury}, \citenamefont {Itano}, \citenamefont {Wineland},\ and\
  \citenamefont {Bergquist}}]{RosHumSch08}%
  \BibitemOpen
  \bibfield  {author} {\bibinfo {author} {\bibfnamefont {T.}~\bibnamefont
  {Rosenband}}, \bibinfo {author} {\bibfnamefont {D.~B.}\ \bibnamefont {Hume}},
  \bibinfo {author} {\bibfnamefont {P.~O.}\ \bibnamefont {Schmidt}}, \bibinfo
  {author} {\bibfnamefont {C.~W.}\ \bibnamefont {Chou}}, \bibinfo {author}
  {\bibfnamefont {A.}~\bibnamefont {Brusch}}, \bibinfo {author} {\bibfnamefont
  {L.}~\bibnamefont {Lorini}}, \bibinfo {author} {\bibfnamefont {W.~H.}\
  \bibnamefont {Oskay}}, \bibinfo {author} {\bibfnamefont {R.~E.}\ \bibnamefont
  {Drullinger}}, \bibinfo {author} {\bibfnamefont {T.~M.}\ \bibnamefont
  {Fortier}}, \bibinfo {author} {\bibfnamefont {J.~E.}\ \bibnamefont
  {Stalnaker}}, \bibinfo {author} {\bibfnamefont {S.~A.}\ \bibnamefont
  {Diddams}}, \bibinfo {author} {\bibfnamefont {W.~C.}\ \bibnamefont {Swann}},
  \bibinfo {author} {\bibfnamefont {N.~R.}\ \bibnamefont {Newbury}}, \bibinfo
  {author} {\bibfnamefont {W.~M.}\ \bibnamefont {Itano}}, \bibinfo {author}
  {\bibfnamefont {D.~J.}\ \bibnamefont {Wineland}}, \ and\ \bibinfo {author}
  {\bibfnamefont {J.~C.}\ \bibnamefont {Bergquist}},\ }\href {\doibase
  10.1126/science.1154622} {\bibfield  {journal} {\bibinfo  {journal}
  {Science}\ }\textbf {\bibinfo {volume} {319}},\ \bibinfo {pages} {1808}
  (\bibinfo {year} {2008})}\BibitemShut {NoStop}%
\bibitem [{\citenamefont {Madej}\ \emph {et~al.}(2012)\citenamefont {Madej},
  \citenamefont {Dub\'e}, \citenamefont {Zhou}, \citenamefont {Bernard},\ and\
  \citenamefont {Gertsvolf}}]{MadDubZho12}%
  \BibitemOpen
  \bibfield  {author} {\bibinfo {author} {\bibfnamefont {A.~A.}\ \bibnamefont
  {Madej}}, \bibinfo {author} {\bibfnamefont {P.}~\bibnamefont {Dub\'e}},
  \bibinfo {author} {\bibfnamefont {Z.}~\bibnamefont {Zhou}}, \bibinfo {author}
  {\bibfnamefont {J.~E.}\ \bibnamefont {Bernard}}, \ and\ \bibinfo {author}
  {\bibfnamefont {M.}~\bibnamefont {Gertsvolf}},\ }\href {\doibase
  10.1103/PhysRevLett.109.203002} {\bibfield  {journal} {\bibinfo  {journal}
  {Phys. Rev. Lett.}\ }\textbf {\bibinfo {volume} {109}},\ \bibinfo {pages}
  {203002} (\bibinfo {year} {2012})}\BibitemShut {NoStop}%
\bibitem [{\citenamefont {Barwood}\ \emph {et~al.}(2014)\citenamefont
  {Barwood}, \citenamefont {Huang}, \citenamefont {Klein}, \citenamefont
  {Johnson}, \citenamefont {King}, \citenamefont {Margolis}, \citenamefont
  {Szymaniec},\ and\ \citenamefont {Gill}}]{BarHuaKle14}%
  \BibitemOpen
  \bibfield  {author} {\bibinfo {author} {\bibfnamefont {G.~P.}\ \bibnamefont
  {Barwood}}, \bibinfo {author} {\bibfnamefont {G.}~\bibnamefont {Huang}},
  \bibinfo {author} {\bibfnamefont {H.~A.}\ \bibnamefont {Klein}}, \bibinfo
  {author} {\bibfnamefont {L.~A.~M.}\ \bibnamefont {Johnson}}, \bibinfo
  {author} {\bibfnamefont {S.~A.}\ \bibnamefont {King}}, \bibinfo {author}
  {\bibfnamefont {H.~S.}\ \bibnamefont {Margolis}}, \bibinfo {author}
  {\bibfnamefont {K.}~\bibnamefont {Szymaniec}}, \ and\ \bibinfo {author}
  {\bibfnamefont {P.}~\bibnamefont {Gill}},\ }\href {\doibase
  10.1103/PhysRevA.89.050501} {\bibfield  {journal} {\bibinfo  {journal} {Phys.
  Rev. A}\ }\textbf {\bibinfo {volume} {89}},\ \bibinfo {pages} {050501}
  (\bibinfo {year} {2014})}\BibitemShut {NoStop}%
\bibitem [{\citenamefont {Godun}\ \emph {et~al.}(2014)\citenamefont {Godun},
  \citenamefont {Nisbet-Jones}, \citenamefont {Jones}, \citenamefont {King},
  \citenamefont {Johnson}, \citenamefont {Margolis}, \citenamefont {Szymaniec},
  \citenamefont {Lea}, \citenamefont {Bongs},\ and\ \citenamefont
  {Gill}}]{GodNisJon14}%
  \BibitemOpen
  \bibfield  {author} {\bibinfo {author} {\bibfnamefont {R.~M.}\ \bibnamefont
  {Godun}}, \bibinfo {author} {\bibfnamefont {P.~B.~R.}\ \bibnamefont
  {Nisbet-Jones}}, \bibinfo {author} {\bibfnamefont {J.~M.}\ \bibnamefont
  {Jones}}, \bibinfo {author} {\bibfnamefont {S.~A.}\ \bibnamefont {King}},
  \bibinfo {author} {\bibfnamefont {L.~A.~M.}\ \bibnamefont {Johnson}},
  \bibinfo {author} {\bibfnamefont {H.~S.}\ \bibnamefont {Margolis}}, \bibinfo
  {author} {\bibfnamefont {K.}~\bibnamefont {Szymaniec}}, \bibinfo {author}
  {\bibfnamefont {S.~N.}\ \bibnamefont {Lea}}, \bibinfo {author} {\bibfnamefont
  {K.}~\bibnamefont {Bongs}}, \ and\ \bibinfo {author} {\bibfnamefont
  {P.}~\bibnamefont {Gill}},\ }\href {\doibase 10.1103/PhysRevLett.113.210801}
  {\bibfield  {journal} {\bibinfo  {journal} {Phys. Rev. Lett.}\ }\textbf
  {\bibinfo {volume} {113}},\ \bibinfo {pages} {210801} (\bibinfo {year}
  {2014})}\BibitemShut {NoStop}%
\bibitem [{\citenamefont {Huntemann}\ \emph {et~al.}(2016)\citenamefont
  {Huntemann}, \citenamefont {Sanner}, \citenamefont {Lipphardt}, \citenamefont
  {Tamm},\ and\ \citenamefont {Peik}}]{HunSanLip16}%
  \BibitemOpen
  \bibfield  {author} {\bibinfo {author} {\bibfnamefont {N.}~\bibnamefont
  {Huntemann}}, \bibinfo {author} {\bibfnamefont {C.}~\bibnamefont {Sanner}},
  \bibinfo {author} {\bibfnamefont {B.}~\bibnamefont {Lipphardt}}, \bibinfo
  {author} {\bibfnamefont {C.}~\bibnamefont {Tamm}}, \ and\ \bibinfo {author}
  {\bibfnamefont {E.}~\bibnamefont {Peik}},\ }\href {\doibase
  10.1103/PhysRevLett.116.063001} {\bibfield  {journal} {\bibinfo  {journal}
  {Phys. Rev. Lett.}\ }\textbf {\bibinfo {volume} {116}},\ \bibinfo {pages}
  {063001} (\bibinfo {year} {2016})}\BibitemShut {NoStop}%
\bibitem [{\citenamefont {Itano}(2000)}]{Ita00}%
  \BibitemOpen
  \bibfield  {author} {\bibinfo {author} {\bibfnamefont {W.~M.}\ \bibnamefont
  {Itano}},\ }\href@noop {} {\bibfield  {journal} {\bibinfo  {journal} {J. Res.
  Natl. Inst. Stand. Technol.}\ }\textbf {\bibinfo {volume} {105}},\ \bibinfo
  {pages} {829} (\bibinfo {year} {2000})}\BibitemShut {NoStop}%
\bibitem [{\citenamefont {Dub\'e}\ \emph {et~al.}(2005)\citenamefont {Dub\'e},
  \citenamefont {Madej}, \citenamefont {Bernard}, \citenamefont {Marmet},
  \citenamefont {Boulanger},\ and\ \citenamefont {Cundy}}]{DubMadBer05}%
  \BibitemOpen
  \bibfield  {author} {\bibinfo {author} {\bibfnamefont {P.}~\bibnamefont
  {Dub\'e}}, \bibinfo {author} {\bibfnamefont {A.~A.}\ \bibnamefont {Madej}},
  \bibinfo {author} {\bibfnamefont {J.~E.}\ \bibnamefont {Bernard}}, \bibinfo
  {author} {\bibfnamefont {L.}~\bibnamefont {Marmet}}, \bibinfo {author}
  {\bibfnamefont {J.-S.}\ \bibnamefont {Boulanger}}, \ and\ \bibinfo {author}
  {\bibfnamefont {S.}~\bibnamefont {Cundy}},\ }\href {\doibase
  10.1103/PhysRevLett.95.033001} {\bibfield  {journal} {\bibinfo  {journal}
  {Phys. Rev. Lett.}\ }\textbf {\bibinfo {volume} {95}},\ \bibinfo {pages}
  {033001} (\bibinfo {year} {2005})}\BibitemShut {NoStop}%
\bibitem [{\citenamefont {Chou}\ \emph {et~al.}(2010)\citenamefont {Chou},
  \citenamefont {Hume}, \citenamefont {Koelemeij}, \citenamefont {Wineland},\
  and\ \citenamefont {Rosenband}}]{ChoHumKoe10}%
  \BibitemOpen
  \bibfield  {author} {\bibinfo {author} {\bibfnamefont {C.~W.}\ \bibnamefont
  {Chou}}, \bibinfo {author} {\bibfnamefont {D.~B.}\ \bibnamefont {Hume}},
  \bibinfo {author} {\bibfnamefont {J.~C.~J.}\ \bibnamefont {Koelemeij}},
  \bibinfo {author} {\bibfnamefont {D.~J.}\ \bibnamefont {Wineland}}, \ and\
  \bibinfo {author} {\bibfnamefont {T.}~\bibnamefont {Rosenband}},\ }\href
  {\doibase 10.1103/PhysRevLett.104.070802} {\bibfield  {journal} {\bibinfo
  {journal} {Phys. Rev. Lett.}\ }\textbf {\bibinfo {volume} {104}},\ \bibinfo
  {pages} {070802} (\bibinfo {year} {2010})}\BibitemShut {NoStop}%
\bibitem [{\citenamefont {Itano}\ \emph {et~al.}(2007)\citenamefont {Itano},
  \citenamefont {Bergquist}, \citenamefont {Brusch}, \citenamefont {Diddams},
  \citenamefont {Fortier}, \citenamefont {Heavner}, \citenamefont {Hollberg},
  \citenamefont {Hume}, \citenamefont {Jefferts}, \citenamefont {Lorini},
  \citenamefont {Parker}, \citenamefont {Rosenband},\ and\ \citenamefont
  {Stalnaker}}]{ItaBerBru07}%
  \BibitemOpen
  \bibfield  {author} {\bibinfo {author} {\bibfnamefont {W.~M.}\ \bibnamefont
  {Itano}}, \bibinfo {author} {\bibfnamefont {J.~C.}\ \bibnamefont
  {Bergquist}}, \bibinfo {author} {\bibfnamefont {A.}~\bibnamefont {Brusch}},
  \bibinfo {author} {\bibfnamefont {S.~A.}\ \bibnamefont {Diddams}}, \bibinfo
  {author} {\bibfnamefont {T.~M.}\ \bibnamefont {Fortier}}, \bibinfo {author}
  {\bibfnamefont {T.~P.}\ \bibnamefont {Heavner}}, \bibinfo {author}
  {\bibfnamefont {L.}~\bibnamefont {Hollberg}}, \bibinfo {author}
  {\bibfnamefont {D.~B.}\ \bibnamefont {Hume}}, \bibinfo {author}
  {\bibfnamefont {S.~R.}\ \bibnamefont {Jefferts}}, \bibinfo {author}
  {\bibfnamefont {L.}~\bibnamefont {Lorini}}, \bibinfo {author} {\bibfnamefont
  {T.~E.}\ \bibnamefont {Parker}}, \bibinfo {author} {\bibfnamefont
  {T.}~\bibnamefont {Rosenband}}, \ and\ \bibinfo {author} {\bibfnamefont
  {J.~E.}\ \bibnamefont {Stalnaker}},\ }\href@noop {} {\bibfield  {journal}
  {\bibinfo  {journal} {Proc. of SPIE}\ }\textbf {\bibinfo {volume} {6673}},\
  \bibinfo {pages} {667303} (\bibinfo {year} {2007})}\BibitemShut {NoStop}%
\bibitem [{\citenamefont {Chen}\ \emph {et~al.}()\citenamefont {Chen},
  \citenamefont {Brewer}, \citenamefont {Hume}, \citenamefont {Chou},
  \citenamefont {Wineland},\ and\ \citenamefont {Leibrandt}}]{CheBreHum16}%
  \BibitemOpen
  \bibfield  {author} {\bibinfo {author} {\bibfnamefont {J.-S.}\ \bibnamefont
  {Chen}}, \bibinfo {author} {\bibfnamefont {S.~M.}\ \bibnamefont {Brewer}},
  \bibinfo {author} {\bibfnamefont {D.~B.}\ \bibnamefont {Hume}}, \bibinfo
  {author} {\bibfnamefont {C.~W.}\ \bibnamefont {Chou}}, \bibinfo {author}
  {\bibfnamefont {D.~J.}\ \bibnamefont {Wineland}}, \ and\ \bibinfo {author}
  {\bibfnamefont {D.~R.}\ \bibnamefont {Leibrandt}},\ }\href@noop {} {}\bibinfo
  {note} {{arXiv:1608.05047 (2016)}}\BibitemShut {NoStop}%
\bibitem [{\citenamefont {Herschbach}\ \emph {et~al.}(2012)\citenamefont
  {Herschbach}, \citenamefont {Pyka}, \citenamefont {Keller},\ and\
  \citenamefont {Mehlst{\"a}ubler}}]{HerPykKel12}%
  \BibitemOpen
  \bibfield  {author} {\bibinfo {author} {\bibfnamefont {N.}~\bibnamefont
  {Herschbach}}, \bibinfo {author} {\bibfnamefont {K.}~\bibnamefont {Pyka}},
  \bibinfo {author} {\bibfnamefont {J.}~\bibnamefont {Keller}}, \ and\ \bibinfo
  {author} {\bibfnamefont {T.~E.}\ \bibnamefont {Mehlst{\"a}ubler}},\ }\href
  {\doibase 10.1007/s00340-011-4790-y} {\bibfield  {journal} {\bibinfo
  {journal} {App. Phys. B}\ }\textbf {\bibinfo {volume} {107}},\ \bibinfo
  {pages} {891} (\bibinfo {year} {2012})}\BibitemShut {NoStop}%
\bibitem [{\citenamefont {Derevianko}(2016)}]{Der16}%
  \BibitemOpen
  \bibfield  {author} {\bibinfo {author} {\bibfnamefont {A.}~\bibnamefont
  {Derevianko}},\ }\href {\doibase 10.1103/PhysRevA.93.012503} {\bibfield
  {journal} {\bibinfo  {journal} {Phys. Rev. A}\ }\textbf {\bibinfo {volume}
  {93}},\ \bibinfo {pages} {012503} (\bibinfo {year} {2016})}\BibitemShut
  {NoStop}%
\bibitem [{\citenamefont {Varshalovich}\ \emph {et~al.}(1988)\citenamefont
  {Varshalovich}, \citenamefont {Moskalev},\ and\ \citenamefont
  {Khersonskii}}]{VarMosKhe88}%
  \BibitemOpen
  \bibfield  {author} {\bibinfo {author} {\bibfnamefont {D.~A.}\ \bibnamefont
  {Varshalovich}}, \bibinfo {author} {\bibfnamefont {A.~N.}\ \bibnamefont
  {Moskalev}}, \ and\ \bibinfo {author} {\bibfnamefont {V.~K.}\ \bibnamefont
  {Khersonskii}},\ }\href@noop {} {\emph {\bibinfo {title} {Quantum Theory of
  Angular Momentum}}}\ (\bibinfo  {publisher} {World Scientific},\ \bibinfo
  {address} {Singapore},\ \bibinfo {year} {1988})\BibitemShut {NoStop}%
\bibitem [{\citenamefont {Dzuba}\ \emph {et~al.}(1996)\citenamefont {Dzuba},
  \citenamefont {Flambaum},\ and\ \citenamefont {Kozlov}}]{DzuFlaKoz96}%
  \BibitemOpen
  \bibfield  {author} {\bibinfo {author} {\bibfnamefont {V.~A.}\ \bibnamefont
  {Dzuba}}, \bibinfo {author} {\bibfnamefont {V.~V.}\ \bibnamefont {Flambaum}},
  \ and\ \bibinfo {author} {\bibfnamefont {M.~G.}\ \bibnamefont {Kozlov}},\
  }\href {\doibase 10.1103/PhysRevA.54.3948} {\bibfield  {journal} {\bibinfo
  {journal} {Phys. Rev. A}\ }\textbf {\bibinfo {volume} {54}},\ \bibinfo
  {pages} {3948} (\bibinfo {year} {1996})}\BibitemShut {NoStop}%
\bibitem [{\citenamefont {Savukov}\ and\ \citenamefont
  {Johnson}(2002)}]{SavJoh02}%
  \BibitemOpen
  \bibfield  {author} {\bibinfo {author} {\bibfnamefont {I.~M.}\ \bibnamefont
  {Savukov}}\ and\ \bibinfo {author} {\bibfnamefont {W.~R.}\ \bibnamefont
  {Johnson}},\ }\href {\doibase 10.1103/PhysRevA.65.042503} {\bibfield
  {journal} {\bibinfo  {journal} {Phys. Rev. A}\ }\textbf {\bibinfo {volume}
  {65}},\ \bibinfo {pages} {042503} (\bibinfo {year} {2002})}\BibitemShut
  {NoStop}%
\bibitem [{\citenamefont {Beloy}\ \emph {et~al.}(2008)\citenamefont {Beloy},
  \citenamefont {Derevianko},\ and\ \citenamefont {Johnson}}]{BelDerJoh08}%
  \BibitemOpen
  \bibfield  {author} {\bibinfo {author} {\bibfnamefont {K.}~\bibnamefont
  {Beloy}}, \bibinfo {author} {\bibfnamefont {A.}~\bibnamefont {Derevianko}}, \
  and\ \bibinfo {author} {\bibfnamefont {W.~R.}\ \bibnamefont {Johnson}},\
  }\href {\doibase 10.1103/PhysRevA.77.012512} {\bibfield  {journal} {\bibinfo
  {journal} {Phys. Rev. A}\ }\textbf {\bibinfo {volume} {77}},\ \bibinfo
  {pages} {012512} (\bibinfo {year} {2008})}\BibitemShut {NoStop}%
\bibitem [{\citenamefont {Dzuba}\ and\ \citenamefont
  {Ginges}(2006)}]{DzuGin06}%
  \BibitemOpen
  \bibfield  {author} {\bibinfo {author} {\bibfnamefont {V.~A.}\ \bibnamefont
  {Dzuba}}\ and\ \bibinfo {author} {\bibfnamefont {J.~S.~M.}\ \bibnamefont
  {Ginges}},\ }\href {\doibase 10.1103/PhysRevA.73.032503} {\bibfield
  {journal} {\bibinfo  {journal} {Phys. Rev. A}\ }\textbf {\bibinfo {volume}
  {73}},\ \bibinfo {pages} {032503} (\bibinfo {year} {2006})}\BibitemShut
  {NoStop}%
\bibitem [{\citenamefont {Lurio}(1962)}]{Lur62}%
  \BibitemOpen
  \bibfield  {author} {\bibinfo {author} {\bibfnamefont {A.}~\bibnamefont
  {Lurio}},\ }\href {\doibase 10.1103/PhysRev.126.1768} {\bibfield  {journal}
  {\bibinfo  {journal} {Phys. Rev.}\ }\textbf {\bibinfo {volume} {126}},\
  \bibinfo {pages} {1768} (\bibinfo {year} {1962})}\BibitemShut {NoStop}%
\bibitem [{\citenamefont {Stone}(2005)}]{Sto05}%
  \BibitemOpen
  \bibfield  {author} {\bibinfo {author} {\bibfnamefont {N.~J.}\ \bibnamefont
  {Stone}},\ }\href {\doibase http://dx.doi.org/10.1016/j.adt.2005.04.001}
  {\bibfield  {journal} {\bibinfo  {journal} {At. Dat. Nucl. Dat. Tab.}\
  }\textbf {\bibinfo {volume} {90}},\ \bibinfo {pages} {75 } (\bibinfo {year}
  {2005})}\BibitemShut {NoStop}%
\bibitem [{\citenamefont {Ralchenko}\ \emph {et~al.}(2016)\citenamefont
  {Ralchenko}, \citenamefont {Kramida}, \citenamefont {Reader},\ and\
  \citenamefont {{NIST ASD Team}}}]{NIST_ASD}%
  \BibitemOpen
  \bibfield  {author} {\bibinfo {author} {\bibfnamefont {Y.}~\bibnamefont
  {Ralchenko}}, \bibinfo {author} {\bibfnamefont {A.~E.}\ \bibnamefont
  {Kramida}}, \bibinfo {author} {\bibfnamefont {J.}~\bibnamefont {Reader}}, \
  and\ \bibinfo {author} {\bibnamefont {{NIST ASD Team}}},\ }\href@noop {}
  {\enquote {\bibinfo {title} {{NIST Atomic Spectra Database (version 5.4)}},}\
  } (\bibinfo {year} {2016}),\ \Eprint
  {http://arxiv.org/abs/http://physics.nist.gov/asd}
  {http://physics.nist.gov/asd} \BibitemShut {NoStop}%
\bibitem [{\citenamefont {Larkins}\ and\ \citenamefont
  {Hannaford}(1993)}]{LarHan93}%
  \BibitemOpen
  \bibfield  {author} {\bibinfo {author} {\bibfnamefont {P.~L.}\ \bibnamefont
  {Larkins}}\ and\ \bibinfo {author} {\bibfnamefont {P.}~\bibnamefont
  {Hannaford}},\ }\href {\doibase 10.1007/BF01437462} {\bibfield  {journal}
  {\bibinfo  {journal} {Z. Phys. D}\ }\textbf {\bibinfo {volume} {27}},\
  \bibinfo {pages} {313} (\bibinfo {year} {1993})}\BibitemShut {NoStop}%
\bibitem [{aun()}]{aunote}%
  \BibitemOpen
  \href@noop {} {}\bibinfo {note} {By definition, $e$, $\hbar$, and $m_e$
  (electron mass) are unity in atomic units. By consequence $a_B$ is also unity
  in atomic units. We reiterate our choice of Gaussian electromagnetic
  expressions.}\BibitemShut {Stop}%
\bibitem [{\citenamefont {Raizen}\ \emph {et~al.}(1992)\citenamefont {Raizen},
  \citenamefont {Gilligan}, \citenamefont {Bergquist}, \citenamefont {Itano},\
  and\ \citenamefont {Wineland}}]{RaiGilBer92}%
  \BibitemOpen
  \bibfield  {author} {\bibinfo {author} {\bibfnamefont {M.~G.}\ \bibnamefont
  {Raizen}}, \bibinfo {author} {\bibfnamefont {J.~M.}\ \bibnamefont
  {Gilligan}}, \bibinfo {author} {\bibfnamefont {J.~C.}\ \bibnamefont
  {Bergquist}}, \bibinfo {author} {\bibfnamefont {W.~M.}\ \bibnamefont
  {Itano}}, \ and\ \bibinfo {author} {\bibfnamefont {D.~J.}\ \bibnamefont
  {Wineland}},\ }\href {\doibase 10.1103/PhysRevA.45.6493} {\bibfield
  {journal} {\bibinfo  {journal} {Phys. Rev. A}\ }\textbf {\bibinfo {volume}
  {45}},\ \bibinfo {pages} {6493} (\bibinfo {year} {1992})}\BibitemShut
  {NoStop}%
\bibitem [{\citenamefont {W\"ubbena}\ \emph {et~al.}(2012)\citenamefont
  {W\"ubbena}, \citenamefont {Amairi}, \citenamefont {Mandel},\ and\
  \citenamefont {Schmidt}}]{WubAmaMan12}%
  \BibitemOpen
  \bibfield  {author} {\bibinfo {author} {\bibfnamefont {J.~B.}\ \bibnamefont
  {W\"ubbena}}, \bibinfo {author} {\bibfnamefont {S.}~\bibnamefont {Amairi}},
  \bibinfo {author} {\bibfnamefont {O.}~\bibnamefont {Mandel}}, \ and\ \bibinfo
  {author} {\bibfnamefont {P.~O.}\ \bibnamefont {Schmidt}},\ }\href {\doibase
  10.1103/PhysRevA.85.043412} {\bibfield  {journal} {\bibinfo  {journal} {Phys.
  Rev. A}\ }\textbf {\bibinfo {volume} {85}},\ \bibinfo {pages} {043412}
  (\bibinfo {year} {2012})}\BibitemShut {NoStop}%
\bibitem [{\citenamefont {Pyka}\ \emph {et~al.}(2014)\citenamefont {Pyka},
  \citenamefont {Herschbach}, \citenamefont {Keller},\ and\ \citenamefont
  {Mehlst{\"a}ubler}}]{PykHerKel14}%
  \BibitemOpen
  \bibfield  {author} {\bibinfo {author} {\bibfnamefont {K.}~\bibnamefont
  {Pyka}}, \bibinfo {author} {\bibfnamefont {N.}~\bibnamefont {Herschbach}},
  \bibinfo {author} {\bibfnamefont {J.}~\bibnamefont {Keller}}, \ and\ \bibinfo
  {author} {\bibfnamefont {T.~E.}\ \bibnamefont {Mehlst{\"a}ubler}},\ }\href
  {\doibase 10.1007/s00340-013-5580-5} {\bibfield  {journal} {\bibinfo
  {journal} {Appl. Phys. B}\ }\textbf {\bibinfo {volume} {114}},\ \bibinfo
  {pages} {231} (\bibinfo {year} {2014})}\BibitemShut {NoStop}%
\bibitem [{\citenamefont {Keller}(2015)}]{Kel15}%
  \BibitemOpen
  \bibfield  {author} {\bibinfo {author} {\bibfnamefont {J.}~\bibnamefont
  {Keller}},\ }\emph {\bibinfo {title} {Spectroscopic characterization of ion
  motion for an optical clock based on Coulomb crystals}},\ \href@noop {}
  {Ph.D. thesis},\ \bibinfo  {school} {Gottfried Wilhelm Leibniz Universit\"at
  Hannover} (\bibinfo {year} {2015})\BibitemShut {NoStop}%
\bibitem [{\citenamefont {Becker}\ \emph {et~al.}(2001)\citenamefont {Becker},
  \citenamefont {Zanthier}, \citenamefont {Nevsky}, \citenamefont {Schwedes},
  \citenamefont {Skvortsov}, \citenamefont {Walther},\ and\ \citenamefont
  {Peik}}]{BecZanNev01}%
  \BibitemOpen
  \bibfield  {author} {\bibinfo {author} {\bibfnamefont {T.}~\bibnamefont
  {Becker}}, \bibinfo {author} {\bibfnamefont {J.~V.}\ \bibnamefont
  {Zanthier}}, \bibinfo {author} {\bibfnamefont {A.~Y.}\ \bibnamefont
  {Nevsky}}, \bibinfo {author} {\bibfnamefont {C.}~\bibnamefont {Schwedes}},
  \bibinfo {author} {\bibfnamefont {M.~N.}\ \bibnamefont {Skvortsov}}, \bibinfo
  {author} {\bibfnamefont {H.}~\bibnamefont {Walther}}, \ and\ \bibinfo
  {author} {\bibfnamefont {E.}~\bibnamefont {Peik}},\ }\href {\doibase
  10.1103/PhysRevA.63.051802} {\bibfield  {journal} {\bibinfo  {journal} {Phys.
  Rev. A}\ }\textbf {\bibinfo {volume} {63}},\ \bibinfo {pages} {051802(R)}
  (\bibinfo {year} {2001})}\BibitemShut {NoStop}%
\bibitem [{\citenamefont {Shaniv}\ \emph {et~al.}(2016)\citenamefont {Shaniv},
  \citenamefont {Akerman},\ and\ \citenamefont {Ozeri}}]{ShaAkeOze16}%
  \BibitemOpen
  \bibfield  {author} {\bibinfo {author} {\bibfnamefont {R.}~\bibnamefont
  {Shaniv}}, \bibinfo {author} {\bibfnamefont {N.}~\bibnamefont {Akerman}}, \
  and\ \bibinfo {author} {\bibfnamefont {R.}~\bibnamefont {Ozeri}},\ }\href
  {\doibase 10.1103/PhysRevLett.116.140801} {\bibfield  {journal} {\bibinfo
  {journal} {Phys. Rev. Lett.}\ }\textbf {\bibinfo {volume} {116}},\ \bibinfo
  {pages} {140801} (\bibinfo {year} {2016})}\BibitemShut {NoStop}%
\end{thebibliography}

%

\end{document}